\documentclass[11pt,epsf]{article}
\DeclareMathAlphabet{\scr}{U}{rsfs}{m}{n}
\usepackage{latexsym,color}
\usepackage{epsfig}
\usepackage[mathscr]{eucal}
\usepackage{amsfonts}
\usepackage{amscd}
\usepackage{cite}
\usepackage{array}
\usepackage{amssymb}
\usepackage{colordvi}
\usepackage[centertags]{amsmath}
\usepackage{enumerate}
\usepackage{graphicx}
\usepackage{booktabs}
\usepackage{theorem}
\usepackage[footnotesize]{caption}
\usepackage{soul}

\newcommand{\lsim}{\raisebox{-0.13cm}{~\shortstack{$<$ \\[-0.07cm] $\sim$}}~}

\newcommand{\newc}{\newcommand}
\newc{\be}{\begin{equation}}
\newc{\ee}{\end{equation}}
\newc{\bea}{\begin{eqnarray}}
\newc{\eea}{\end{eqnarray}}
\newc{\ol}{\overline}
\newc{\wt}{\widetilde}
\newc{\bs}{\boldsymbol}
\newc{\m}{\mathcal}
\newc{\la}{\langle}
\newc{\ra}{\rangle}

\newcommand{\beq}{\begin{eqnarray}}
\newcommand{\eeq}{\end{eqnarray}}
\newcommand{\s}{\smallskip}

\newcommand{\bc}{\begin{center}}
\newcommand{\ec}{\end{center}}

\newcommand{\ba}{\begin{array}}
\newcommand{\ea}{\end{array}}

\begin{document}

\title{
\vspace*{-3cm}
\phantom{h} \hfill\mbox{\small KA-TP-08-2015}\\[-1.1cm]
\phantom{h} \hfill\mbox{\small PSI-PR-15-03}\\[-1.1cm]
\phantom{h} \hfill\mbox{\small RM3-TH-15-5}
\\[1cm]
\textbf{NLO QCD Corrections to Higgs Pair Production including
  Dimension-6 Operators}}

\date{}
\author{
R.~Gr\"ober$^{1\,}$\footnote{E-mail: \texttt{groeber@roma3.infn.it}},
M. M\"{u}hlleitner$^{2\,}$\footnote{E-mail: \texttt{milada.muehlleitner@kit.edu}},
M.~Spira$^{3\,}$\footnote{E-mail: \texttt{michael.spira@psi.ch}} $\;$and
 J.~Streicher$^{2\,}$\footnote{E-mail: \texttt{juraj.streicher@kit.edu}}
\\[9mm]
{\small\it
$^1$ INFN, Sezione di Roma Tre, Via della Vasca Navale 84, I-00146
Roma, Italy}\\[3mm] 
{\small\it
$^2$Institute for Theoretical Physics, Karlsruhe Institute of Technology,} \\
{\small\it 76128 Karlsruhe, Germany}\\[3mm]
{\small\it
$^3$ Paul Scherrer Institute, CH-5323 Villigen PSI, Switzerland}}

\maketitle

\begin{abstract}
\noindent
New Physics that becomes relevant at some high scale $\Lambda$ beyond
the experimental reach, can be described in the effective theory
approach by adding higher-dimensional operators to the Standard Model
(SM) Lagrangian. In Higgs pair production through gluon fusion, which
gives access to the trilinear Higgs self-coupling, this leads not only
to modifications of the SM couplings but also induces novel couplings
not present in the SM. For a proper prediction of the cross section,
higher order QCD corrections that are important for this process, have to be taken
into account. The various higher-dimensional contributions are affected
differently by the QCD corrections. In this paper, we provide the
next-to-leading order (NLO) QCD corrections to Higgs pair production
including dimension-6 operators in the limit of large top quark
masses. Depending on the dimension-6 coefficients entering the
Lagrangian, the new operators affect the relative NLO QCD corrections
by several per cent, while modifying the cross section by up to an
order of magnitude.
\end{abstract}
\thispagestyle{empty}
\vfill
\newpage
\setcounter{page}{1}

\section{Introduction}
With the discovery of the Higgs boson \cite{:2012gk,:2012gu}, its role
has developed from the long-sought particle into a tool for exploring beyond the
Standard Model (SM) physics \cite{Englert:2014uua}, possibly paving
the way into new physics (NP) territory. This is the more true, as 
to date we are lacking any direct evidence of physics beyond the SM
(BSM). The Higgs boson itself behaves SM-like with its couplings
to other SM particles being close to the predicted values, in
particular the couplings to gauge bosons. In some NP models, however,
the trilinear Higgs self-coupling can still deviate significantly from
the SM expectations \cite{Azatov:2015oxa}. A means
to describe BSM physics, that is realized at a scale well above the
electroweak symmetry breaking scale, in a rather model-independent way
is given by the Effective Field Theory (EFT) framework. Deviations from the SM are
parametrized by higher-dimensional operators, which lead to
modifications of the Higgs boson couplings to the other SM particles
and to itself. \s 

The trilinear Higgs self-coupling is accessible in double Higgs
production \cite{Dawson:1998py,Djouadi:1999gv,Djouadi:1999rca,thesis},
with the dominant production mechanism at the LHC provided by gluon fusion
\cite{Plehn:1996wb,Baglio:2012np}. The leading order process is
mediated by top and bottom quark triangle and box diagrams. As in single
Higgs production 
\cite{singlenlo,singlenloapprox}, the next-to-leading order (NLO) QCD
corrections to this process are important. They have first 
been obtained in the limit of large top quark masses \cite{Dawson:1998py}. While
this approximation works quite well for single Higgs production, the
uncertainties of the approximation are considerably larger for double
Higgs production and even more in the case of differential
distributions \cite{Baur:2002rb,Gillioz:2012se}. Top quark mass effects have
been analysed in
\cite{Dawson:2012mk,Grigo:2013rya,Frederix:2014hta,Maltoni:2014eza},
and first results towards a fully differential NLO calculation have been presented in
\cite{Frederix:2014hta}. Recently, the
next-to-next-to-leading order (NNLO) QCD corrections have been calculated in
\cite{deFlorian:2013uza,deFlorian:2013jea,Grigo:2014jma}. The authors
of \cite{Shao:2013bz} have performed a soft gluon resummation at
next-to-next-to-leading logarithmic order within the SCET approach. \s

Higher-dimensional operators relevant for Higgs pair production
through gluon fusion have been discussed in
\cite{Azatov:2015oxa,Contino:2012xk,Chen:2014xra,Goertz:2014qta,Edelhaeuser:2015zra}. They lead to contributions that are different
for the triangle and box diagrams mediating the pair
production process. In this work we perform the computation of the NLO QCD
corrections to gluon fusion into Higgs pairs including higher
dimensional operators in the large top mass limit. Our result
allows us to investigate the validity of an approximation applied in previous
works. This approximation relies on the multiplication of the full
leading order (LO) cross section by an overall $K$-factor, given by
the ratio of the SM result for the NLO QCD cross section divided by
the LO cross section, in the large top mass limit.  \s

In the next section \ref{sec:calc} we present the details of our
calculation. This is followed by a numerical analysis in section
\ref{sec:numerical}. In section \ref{sec:concl} we summarize and conclude.

\section{Details of the calculation \label{sec:calc}}
Gluon fusion into Higgs pairs is mediated by top and bottom quark
loops dominantly \cite{Plehn:1996wb}. We compute the NLO QCD corrections in the
heavy top quark limit and we neglect in the following in this
framework the bottom quark loops, which only contribute with less than 1\%
\cite{Dawson:1998py,Baglio:2012np}. \s

If physics beyond the SM appears at some high-scale, NP effects can be
parametrized in a rather model-independent way by introducing
higher-dimensional operators. In case the Higgs boson is embedded in an
$SU(2)$-doublet $H$ the leading BSM effects are
given by dimension-6 operators.\footnote{In certain parameter regions
  dimension-8 operators can become more important than the dimension-6
ones \cite{Azatov:2015oxa}. Since the investigation of the concerned
kinematic regions is challenging we neglect those operators in the
following.} In the Strongly-Interacting-Light Higgs (SILH) basis the
operators relevant for Higgs pair production are given by \cite{silh},
\beq
\Delta {\cal L}_6^{\text{SILH}} &\supset&
\frac{\bar{c}_H}{2v^2} \partial_\mu (H^\dagger H) \partial^\mu
(H^\dagger H) + \frac{\bar{c}_u}{v^2} y_t (H^\dagger H \bar{q}_L H^c
t_R + h.c.) \nonumber \\
&& - \frac{\bar{c}_6}{6 v^2} \frac{3 M_h^2}{v^2} (H^\dagger H)^3
+ \bar{c}_g \frac{g_s^2}{m_W^2} H^\dagger H G_{\mu\nu}^a G^{a\,
  \mu\nu} \;, \label{eq:silh}
\eeq 
where $v$ is the vacuum expectation value $v\approx 246$~GeV,
$M_h=125$~GeV the Higgs boson mass, $m_W$ 
the $W$ boson mass, $y_t$ the top Yukawa coupling constant, $g_s$ the
strong coupling constant and 
$G^a_{\mu\nu}$ the gluon field strength tensor. Note that we neglect
CP-violating effects. An estimate of the size of the coefficients
$\bar{c}_H, \bar{c}_u, \bar{c}_6$ and $\bar{c}_g$ and the most
important experimental bounds can be found in
\cite{Contino:2013kra}. The first three operators in
Eq.~(\ref{eq:silh}) modify the top Yukawa and the trilinear Higgs
self-coupling with respect to the corresponding SM values, while the last operator 
parametrizes effective gluon couplings to one and two Higgs
bosons not mediated by SM quark loops. The second operator furthermore
introduces a novel two-Higgs two-fermion coupling \cite{Grober:2010yv}. \s

In case the $SU(2)_L \times U(1)_Y$ is non-linearly realized and the
physical Higgs boson $h$ is a singlet of the custodial symmetry and not
necessarily part of a weak doublet, the contributions relevant for
our process are summarized by the non-linear Lagrangian \cite{Contino:2010mh}
\beq
\Delta {\cal L}_{\text{non-lin}} \supset - m_t \bar{t}t \left(c_t
  \frac{h}{v} + c_{tt} \frac{h^2}{2 v^2} \right) 
-c_3 \, \frac{1}{6} \left( \frac{3 M_h^2}{v} \right) h^3 + \frac{\alpha_s}{\pi} G^{a\, \mu\nu}
G_{\mu\nu}^a \left( c_g \frac{h}{v} + c_{gg}\frac{h^2}{2 v^2}
\right) \;,
\eeq
with $\alpha_s = g_s^2/(4\pi)$. In contrast to the SILH
parametrization, where the coupling  deviations from the SM are
required to be small, the non-linear Lagrangian is
valid for arbitrary values of the couplings $c_i$. From the SILH
Lagrangian in the unitary gauge and after canonical normalization 
the relations between the SILH coefficients and
the non-linear coefficients $c_i$ can be derived, leading to \cite{Azatov:2015oxa}
\beq
c_t = 1 - \frac{\bar{c}_H}{2} - \bar{c}_u \;, \quad c_{tt} = -
\frac{1}{2} (\bar{c}_H + 3 \bar{c}_u ) \;, \quad c_3 = 1- \frac{3}{2}
\bar{c}_H + \bar{c}_6 \;, \quad c_g = c_{gg} = \bar{c}_g \left(
  \frac{4 \pi}{\alpha_2} \right) \;, \label{eq:nonlincoeff}
\eeq
with $\alpha_2 = \sqrt{2} G_F m_W^2/\pi$ and $G_F$ denoting the Fermi
constant. In the following we will give results for the non-linear
parametrization and defer the SILH case to
Appendix~\ref{sec:silhggfus}. \s 
  
\begin{figure}[h]
\begin{minipage}{.25\textwidth}
\includegraphics{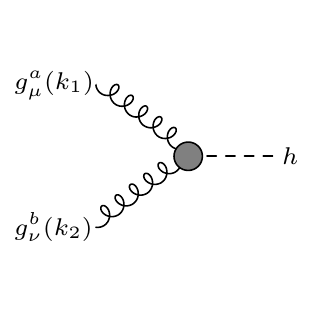}
\end{minipage}
\begin{minipage}{.75\textwidth}
$i \delta^{ab} \frac{\alpha_s}{3\pi v} [k_1^\nu k_2^\mu - (k_1 \cdot
k_2) g^{\mu\nu}] \left[ c_t \left(1+\frac{11}{4} \frac{\alpha_s}{\pi} \right)+
  12 c_{g} \right]$
\end{minipage}\\
\begin{minipage}{.25\textwidth}
\includegraphics{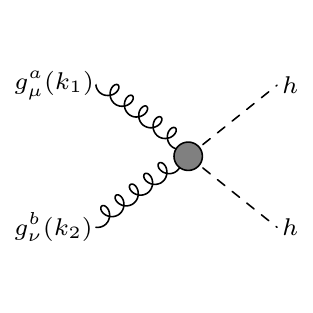}
\end{minipage}
\begin{minipage}{.75\textwidth}
\begin{flushleft}
$i \delta^{ab} \frac{\alpha_s}{3\pi v^2} [k_1^\nu k_2^\mu - (k_1 \cdot
k_2) g^{\mu\nu}] \left[ (c_{tt}-c_t^2) \left(1+\frac{11}{4}
    \frac{\alpha_s}{\pi} \right) + 12 c_{gg}\right]$ \\
\end{flushleft}
\end{minipage}
\caption{Feynman rules for the effective two-gluon couplings to one
  and two Higgs bosons in the heavy quark limit, including NLO QCD 
  corrections. The incoming four-momenta of the gluons are denoted by
  $k_1$ and $k_2$. \label{fig:effvertices}}
\end{figure}

In the low-energy limit of small Higgs four-momentum an effective
Lagrangian valid for light Higgs bosons can be derived for the Higgs
boson interactions. The effective Lagrangian can be
used to compute the QCD corrections to Higgs pair production in the
large top mass limit. From single-Higgs production it is known that
the $K$-factor obtained in this limit approximates the exact value to
better than 5\%, when the full mass dependence is included in the LO
cross section \cite{singlenlo}. The low-energy approach has also
been used to derive the QCD corrections to Higgs pair production
\cite{Dawson:1998py}. Here the uncertainty of 10\% induced in the $K$-factor is
worse than in the single Higgs case
\cite{Dawson:2012mk,Grigo:2013rya,Frederix:2014hta,Maltoni:2014eza}.\footnote{Note,
however, that the application of the LET in minimal composite Higgs
models at LO leads to an even worse approximation than in the
SM \cite{Gillioz:2012se,Grober:2010yv}. Since the NLO corrections
are dominated by soft and collinear gluon effects, the top mass effects on the
$K$-factor can be expected to be of order 10--20\% also for models including
higher-dimensional operators.} The 
effective Lagrangian for the Higgs couplings to gluons and quarks
reads \cite{Dawson:1998py} 
\beq
{\cal L}_{\text{eff}} = \frac{\alpha_s}{\pi} G^{a\mu\nu}
G_{\mu\nu}^a && \hspace*{-0.6cm} \left\{ \frac{h}{v} \left[\frac{c_t}{12} \left( 1 +
      \frac{11}{4} \frac{\alpha_s}{\pi} \right) + c_g
  \right] \right. \nonumber \\
&& \hspace*{-0.6cm} \left. + \frac{h^2}{v^2} \left[
    \frac{-c_t^2+c_{tt}}{24} \left( 1 + 
      \frac{11}{4} \frac{\alpha_s}{\pi} \right) + \frac{c_{gg}}{2}
  \right] \right\}\;.
\eeq
The factor $(1+11/4 \, \alpha_s/\pi)$ emerges from the matching of the effective 
to the full theory at NLO QCD. The Feynman rules for the effective
couplings of two gluons to one and two Higgs bosons,
based on the low-energy theorems \cite{Ellis:1975ap,Kniehl:1995tn},
are given in Fig.~\ref{fig:effvertices}.

\begin{figure}[t]
\begin{minipage}{0.33\textwidth}
 \includegraphics[width=\textwidth]{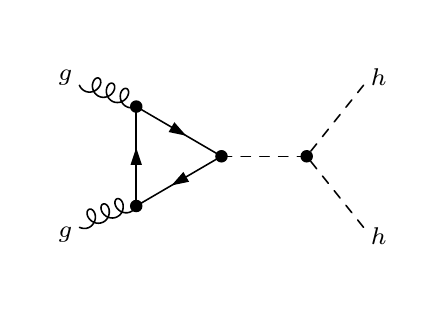}
\end{minipage}
\begin{minipage}{0.33\textwidth}
 \includegraphics[width=\textwidth]{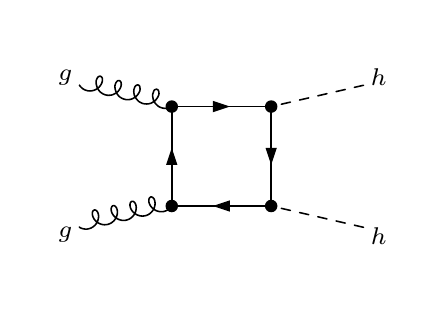}
\end{minipage}
\begin{minipage}{0.33\textwidth}
 \includegraphics[width=\textwidth]{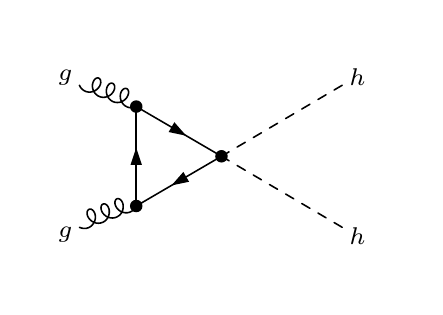}
\end{minipage}\\
 \begin{minipage}{0.49\textwidth}
\begin{flushright}
 \includegraphics[width=0.66\textwidth]{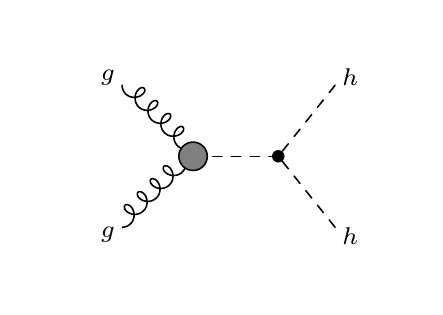}
\end{flushright}
\end{minipage}
\begin{minipage}{0.49\textwidth}
\begin{flushleft}
 \includegraphics[width=0.66\textwidth]{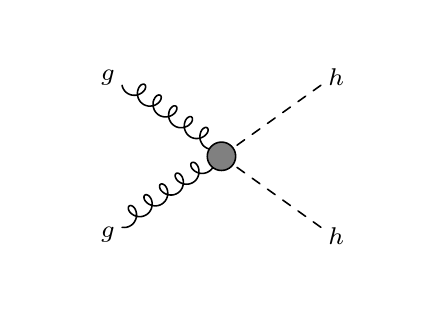}
\end{flushleft}
\end{minipage}

\caption{Generic diagrams contributing to Higgs pair production in
  gluon fusion at LO. \label{fig:lodiagrams}} 
\end{figure}
The generic diagrams contributing to gluon fusion into Higgs pairs at
LO are depicted in Fig.~\ref{fig:lodiagrams}. The LO partonic cross
section can generically be cast into the
form 
\beq
\hat{\sigma}_{\text{LO}} (gg \to hh) = \int_{\hat{t}_-}^{\hat{t}_+} 
d\hat{t} \, \frac{G_F^2 \alpha_s^2(\mu_R)}{256 (2\pi)^3} \,
\left[
\left| C_\Delta F_1 + F_2 \right|^2   + |c_t^2 G_\Box|^2 
 \right] \;, \label{eq:losigma} 
\eeq
where $\mu_R$ denotes the renormalization scale. The Mandelstam
variables are given by 
\beq
\hat{s} &=& Q^2 \;, \qquad 
\hat{t} = M_h^2 - \frac{Q^2(1-\beta \cos \theta)}{2}  \qquad
\mbox{and} \qquad
\hat{u} = M_h^2 - \frac{Q^2(1+\beta \cos \theta)}{2} \;,
\eeq
in terms of the scattering angle $\theta$ in the partonic
center-of-mass (c.m.)~system
with the invariant Higgs pair mass $Q$ and the relative velocity
\beq
\beta = \sqrt{1-\frac{4M_h^2}{Q^2}}\;.
\eeq
The integration limits at $\cos\theta = \pm 1$ are
\beq
\hat{t}_\pm = M_h^2 -\frac{Q^2(1 \mp \beta)}{2} \;.
\eeq
The form factors $F_\Delta$ and $F_\Box$ in $F_1$ and $F_2$
defined as  
\beq
F_1 &=& c_t F_\Delta + \frac{2}{3} c_\Delta \qquad \mbox{and} \qquad
F_2 = c_t^2 F_\Box +c_{tt} F_\Delta - \frac{2}{3} c_\Box \;,
\eeq
contain the full
quark mass dependence and can be found in \cite{Plehn:1996wb}. In
the heavy quark limit the form factors $F_\Delta$, $F_\Box$ and $G_\Box$ approach
\beq
F_\Delta \to \frac{2}{3} \;, \qquad F_\Box \to -\frac{2}{3} \qquad
\mbox{and} \qquad G_\Box = 0 \;,
\eeq
and $F_1$ and $F_2$ simplify to
\beq
F_1^{\text{lim}} &=& \frac{2}{3} \left( c_t + c_\Delta \right) \;, \qquad
F_2^{\text{lim}} = \frac{2}{3} (- c_t^2 + c_{tt} - c_\Box) \;.
\eeq
We have furthermore introduced the abbreviations 
\beq
C_\Delta \equiv \lambda_{hhh} \frac{M_Z^2}{Q^2-M_h^2+i M_h \Gamma_h}
\;, \qquad c_\Delta \equiv 12 c_{g} \qquad
\mbox{and} \qquad c_\Box \equiv - 12 c_{gg} \;,
\eeq
with 
\beq
\lambda_{hhh} = \frac{3 M_h^2 c_3}{M_Z^2} \;.
\eeq
The terms proportional to $c_t$, respectively $c_t^2$ in $F_1$ and
$F_2$ and in front of the form factor $G_\Box$ are the usual SM
contributions including the modifications due to the 
rescaling of the top Yukawa coupling by $c_t$. The contributions
coming with $c_\Delta$ and $c_\Box$ originate from the 
effective two-gluon couplings to one and two Higgs bosons, while
the term involving $c_{tt}$ is due to the novel two-Higgs two-top
quark coupling. \s 

\begin{figure}
\begin{minipage}{0.33\textwidth}
\includegraphics[width=\textwidth]{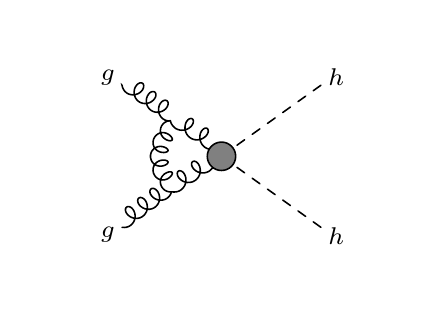}
\end{minipage}
\begin{minipage}{0.33\textwidth}
\includegraphics[width=\textwidth]{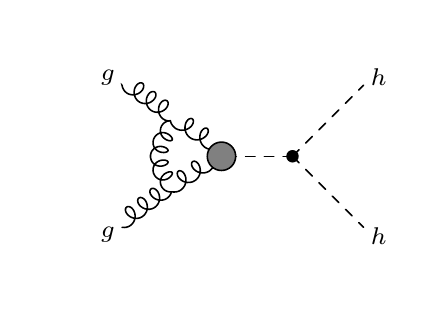}
\end{minipage}
\begin{minipage}{0.33\textwidth}
\includegraphics[width=\textwidth]{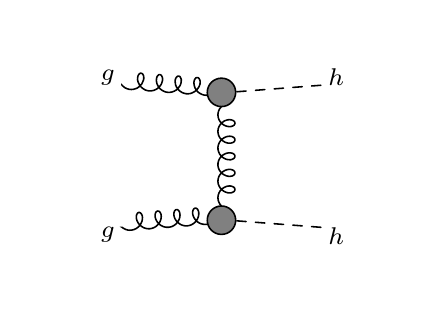}
\end{minipage}\\
\begin{minipage}{0.33\textwidth}
\includegraphics[width=\textwidth]{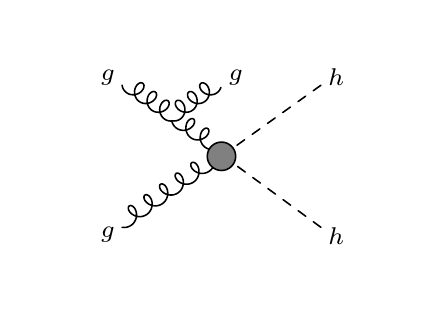}
\end{minipage}
\begin{minipage}{0.33\textwidth}
\includegraphics[width=\textwidth]{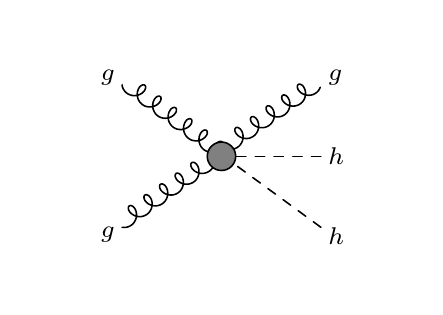}
\end{minipage}
\begin{minipage}{0.33\textwidth}
\includegraphics[width=\textwidth]{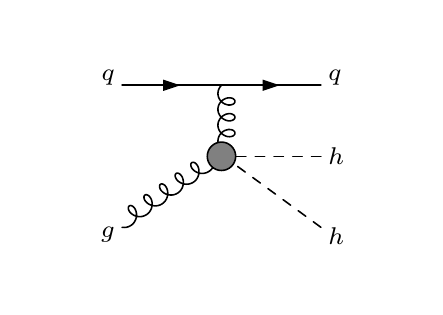}
\end{minipage}
\caption{Sample effective diagrams contributing to the virtual
  (upper) and the real (lower) corrections to gluon fusion into Higgs
  pairs. \label{fig:nlodiags}} 
\end{figure}
We use the effective couplings to compute the NLO QCD corrections to Higgs
pair production. They are composed of the virtual and the real
corrections. Sample diagrams are shown in
Fig.~\ref{fig:nlodiags}. The calculation is performed by applying
dimensional regularization in $d=4-2\epsilon$ dimensions. The
ultraviolet divergences are canceled through the renormalization of
the strong coupling constant in the $\overline{\mbox{MS}}$ scheme with
five active flavours, {\it i.e.}~with the top quark decoupled from the
running of $\alpha_s$. The infrared divergences are canceled by
summing the virtual and the real corrections. The remaining collinear
initial state singularities are absorbed into the NLO parton
densities, which are defined in the $\overline{\mbox{MS}}$ scheme with
five light quark flavours. The finite hadronic NLO cross section can
then be written as
\beq
\sigma_{\text{NLO}} (pp \to hh + X) = \sigma_{\text{LO}} +
\Delta \sigma_{\text{virt}} + \Delta \sigma_{gg} + \Delta \sigma_{gq}
+ \Delta \sigma_{q\bar{q}} \;. \label{eq:nlosigma}
\eeq
We obtain for the individual contributions of Eq.~(\ref{eq:nlosigma})
\beq
\sigma_{\text{LO}} &=& \int_{\tau_0}^1 d\tau \, \frac{d {\cal
    L}^{gg}}{d\tau} \, \hat{\sigma}_{\text{LO}} (Q^2 = \tau s) \\
\Delta \sigma_{\text{virt}} &=& \frac{\alpha_s (\mu_R)}{\pi} \int_{\tau_0}^1
d\tau \, \frac{d {\cal L}^{gg}}{d\tau} \, \hat{\sigma}_{\text{LO}} (Q^2 =
\tau s) \, C \label{eq:contrib1}\\
\Delta \sigma_{gg} &=& \frac{\alpha_s (\mu_R)}{\pi} \int_{\tau_0}^1
d\tau \, \frac{d {\cal L}^{gg}}{d\tau} 
\int_{\tau_0/\tau}^1 \frac{dz}{z} \, \hat{\sigma}_{\text{LO}} (Q^2 =
z\tau s) \left\{ -z P_{gg} (z) \log \frac{\mu_F^2}{\tau s} \right. \nonumber \\
&& \hspace*{2cm} \left. -\frac{11}{2} (1-z)^3 + 6[1+z^4+(1-z)^4] \left(
    \frac{\log (1-z)}{1-z} \right)_+ \right\} \label{eq:contrib2} \\
\Delta \sigma_{gq} &=& \frac{\alpha_s (\mu_R)}{\pi} \int_{\tau_0}^1
d\tau \, \sum_{q,\bar{q}} \frac{d {\cal L}^{gq}}{d\tau} 
\int_{\tau_0/\tau}^1 \frac{dz}{z} \, \hat{\sigma}_{\text{LO}} (Q^2 =
z\tau s) \left\{ - \frac{z}{2} P_{gq} (z) \log \frac{\mu_F^2}{\tau s
    (1-z)^2} \right. \nonumber \\
&& \left. \hspace*{5cm} + \frac{2}{3} z^2 - (1-z)^2
\right\} \label{eq:contrib3} \\
\Delta \sigma_{q\bar{q}} &=& \frac{\alpha_s (\mu_R)}{\pi} \int_{\tau_0}^1
d\tau \, \sum_q \frac{d {\cal L}^{q\bar{q}}}{d\tau} 
\int_{\tau_0/\tau}^1 \frac{dz}{z} \, \hat{\sigma}_{\text{LO}} (Q^2 =
z\tau s) \, \frac{32}{27} (1-z)^3 \;, \label{eq:contrib4}
\eeq
where $s$ denotes the hadronic c.m.~energy and 
\beq
\tau_0 = \frac{4 M_h^2}{s} \;.
\eeq
The Altarelli-Parisi splitting functions are given by \cite{altaparisi},
\beq
P_{gg} (z) &=& 6 \left\{ \left( \frac{1}{1-z} \right)_+ + \frac{1}{z}
  - 2 + z (1-z) \right\} + \frac{33-2N_F}{6} \delta (1-z) \nonumber \\
P_{gq} (z) &=& \frac{4}{3} \frac{1+(1-z)^2}{z} \;,
\eeq
with $N_F=5$ in our case. We denote the factorization scale of the
parton-parton luminosities $d{\cal L}^{ij}/d\tau$ by $\mu_F$. While the
relative real corrections are not affected by the higher-dimensional operators,
the virtual corrections are altered compared to the SM case because of
the overall coupling modifications of the top Yukawa and the trilinear
Higgs self-coupling and due to the additional contributions from the novel
effective vertices. The coefficient $C$ for the virtual corrections
reads
\beq
C &=& \pi^2 + \frac{33-2N_F}{6} \log \frac{\mu_R^2}{Q^2} +
\frac{11}{2} \nonumber \\
&& + \, \mbox{Re} \, \frac{\int_{\hat{t}_-}^{\hat{t}_+} d\hat{t}
   \, \left[ C_\Delta F_1
      +  F_2 \right] \left[ - C_\Delta^* \frac{11}{3} c_\Delta 
     +  \frac{11}{3} c_\Box+c_1(c_t+c_\Delta)^2 \right]
 }{\int_{\hat{t}_-}^{\hat{t_+}} d\hat{t}\, \left[ \left| C_\Delta F_1
      +  F_2 \right|^2 + |c_t^2  G_\Box|^2 \right]}  
\label{eq:ccoeffnew}
\\ && + c_t^2 (c_t + c_\Delta)^2 \, 
\mbox{Re} \, \frac{\int_{\hat{t}_-}^{\hat{t}_+} d\hat{t} \left[ c_2
    \frac{p_T^2}{2 \hat{t} 
      \hat{u}} (Q^2-2M_h^2) G_\Box  \right]}{
\int_{\hat{t}_-}^{\hat{t}_+} d\hat{t} \, \left[\left| C_\Delta F_1
      + F_2 \right|^2 + |c_t^2 G_\Box|^2\right]}  
\;, \nonumber
\eeq
with the squared transverse momentum 
\beq
p_T^2 = \frac{(\hat{t}-M_h^2)(\hat{u}-M_h^2)}{Q^2} - M_h^2
\eeq
and the coefficients 
\beq
c_1 = -c_2 = \frac{4}{9} \;.
\eeq
The third line in Eq.~(\ref{eq:ccoeffnew}) and the terms proportional
to $c_1$ in the second line originate from the third diagram with the two effective
Higgs-two-gluon couplings in Fig.~\ref{fig:nlodiags} (upper), and the
remaining terms are due to the diagrams with gluon
loops in Fig.~\ref{fig:nlodiags} (upper).  
In the derivation of the coefficient $C$ for the virtual corrections we have
kept the full top quark mass dependence in the LO amplitude. The SM result is
recovered for
\beq
c_t \to 1 \;, \quad c_{tt} \to 0 \;, \quad c_3 \to 1 \;, \quad c_\Delta \to
0 \quad \mbox{and} \quad c_\Box \to 0 \;.
\eeq

\section{Numerical Analysis \label{sec:numerical}}
For the numerical analysis we have implemented the LO and NLO Higgs
pair production cross sections including the contributions of dimension-6
operators, as presented in the previous section, in the Fortran program
{\tt HPAIR} \cite{hpair}. We have chosen the c.m.~energy $\sqrt{s}=14$~TeV and for
comparison also the very high energy  
option $\sqrt{s}=100$~TeV. The Higgs boson mass has been set equal to
$M_h=125$~GeV \cite{Aad:2015zhl} and for the top and bottom quark
masses we have chosen $m_t=173.2$~GeV and $m_b=4.75$~GeV,
respectively. We have adopted the MSTW08 \cite{mstw08} parton
densities for the LO\footnote{Note that in the LO cross section we
  have neglected the bottom quark loops. Their effect amounts to less
  than 1\%. This treatment accounts for the fact that we did not 
  introduce a coupling modification factor in the Higgs Yukawa
  coupling to bottom quarks. Additionally it is consistent with the
  application of the heavy top quark limit in the NLO QCD
  corrections.} and NLO cross sections with $\alpha_s 
(M_Z)=0.13939$ at LO and $\alpha_s (M_Z)=0.12018$ at
NLO. \s

\begin{figure}[t!]
\begin{center}
\includegraphics[scale=0.5]{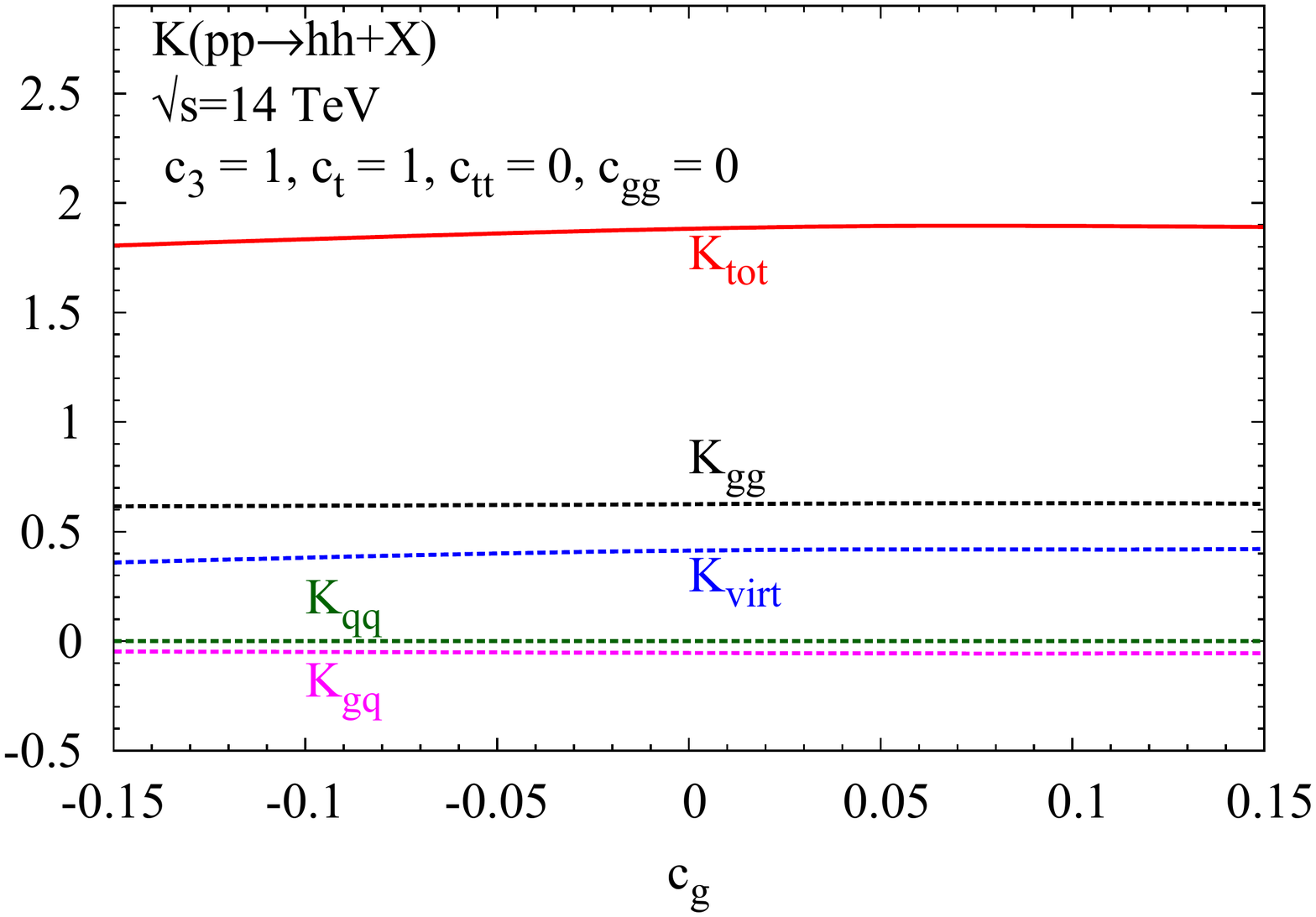} \\[-1.4cm]
\includegraphics[scale=0.5]{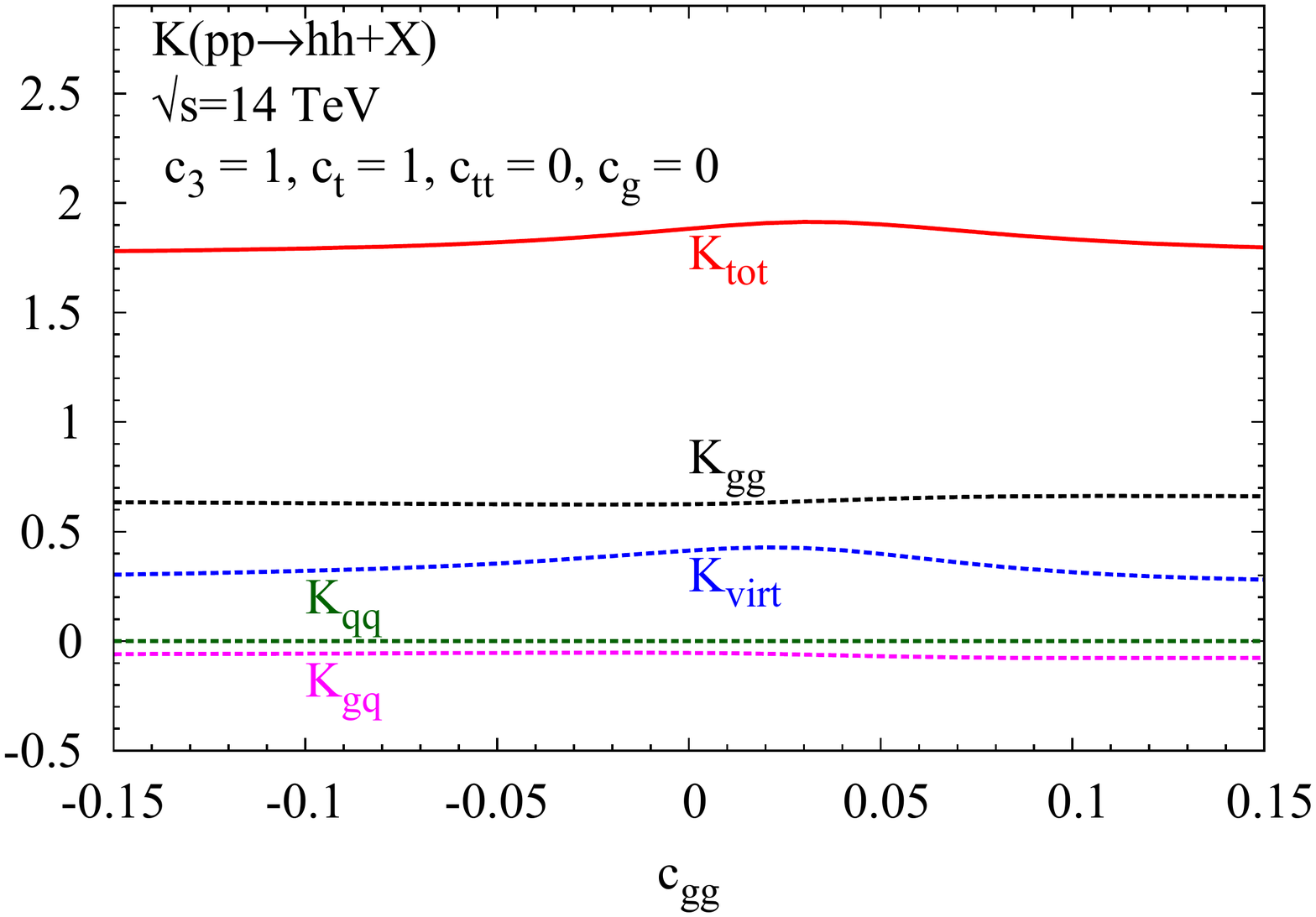}
\vspace*{-0.5cm}
  \caption{$K$-factors of the QCD-corrected gluon fusion cross
    section $\sigma (pp \to hh +X)$ at the LHC with c.m. energy
    $\sqrt{s}=14$~TeV.  The dashed lines show the individual
    contributions of the four terms of the QCD corrections given in 
    Eq.~(\ref{eq:nlosigma}), {\it
      i.e.}~$K_i=\Delta \sigma_i/\sigma_{\text{LO}}$
    ($i=\text{virt},gg, gq, q\bar{q}$). We have set the SM
    values $c_3=c_t=1$, $c_{tt}=0$ and varied $c_g$ with
    $c_{gg}=0$ (upper), respectively, varied $c_{gg}$ with $c_g=0$
    (lower). 
\label{fig:kfactorcgandcgg}}
\end{center}
\end{figure}
In order to study the impact of the new couplings on the QCD
corrections we show in Fig.~\ref{fig:kfactorcgandcgg} the
$K$-factor, defined as the ratio of the NLO
and LO cross sections, $K=\sigma_{\text{NLO}}/\sigma_{\text{LO}}$,
where the parton densities and the strong couplings $\alpha_s$ are
taken at NLO and LO, respectively. 
Deviations with respect to the SM $K$-factor arise in the virtual
corrections due to the second and third line in the formula
Eq.~(\ref{eq:ccoeffnew}) for the coefficient $C$. Additionally, the
real corrections are affected because of the different weights in the
$\tau$ integration due to the modified LO cross section. 
In Fig.~\ref{fig:kfactorcgandcgg}
(upper) we have set all couplings but $c_g$ to their SM values,
$c_3=c_t=1$ and $c_{tt}=c_{gg}=0$. We have varied $c_g$ away from its
SM value $c_g=0$ in the range $-0.15 \le c_g \le 0.15$. For
illustrative purposes we have chosen a rather large range, that goes
beyond current experimental limits obtained under certain assumptions
\cite{Azatov:2015oxa,cmshig009}. In 
the lower plot we have instead set $c_g=0$ and varied $c_{gg}$ away
from its SM value $c_{gg}=0$ in the range $-0.15 \le c_{gg} \le
0.15$ \cite{Azatov:2015oxa}. As can be inferred from
Fig.~\ref{fig:kfactorcgandcgg} the new 
contact interactions $c_g$ and $c_{gg}$ each vary the $K$-factor
between 1.8 and 1.9 in the investigated range. 
Figure~\ref{fig:kfactorcgandcgg} (lower) shows that
the maximal deviation from the SM $K$-factor is obtained for
$c_{gg}=-0.15$, where the $K$-factor practically becomes constant. It
amounts to
\beq
\delta_{\,\text{max}}^{\,c_{gg}} =
\frac{\mbox{max}|K^{c_{gg}}-K^{\text{SM}}|}{K^{\text{SM}}} =  5.4\% \;.
\eeq 
The impact on the total cross section, however, is much larger. Here
we have at NLO 
$\mbox{max}|\sigma^{c_{gg}}-\sigma^{\text{SM}}|/\sigma^{\text{SM}}
= 5.8$. While the effect of higher-dimensional operators on the
$K$-factor is at the level of a few per cent, on the cross section itself it
is enormous. 
In Fig.~\ref{fig:kfactorcgandcgg} (upper) the maximum is less pronounced,
as $c_g$ modifies the coupling of a single Higgs boson to two gluons, which
is attached to the Higgs 
propagator. At the c.m.~energy of $\sqrt{s}=100$~TeV (not shown here)
the $K$-factor is smaller and varies for non-zero $c_{gg}$ in the
range 1.46--1.58. For $c_g$ the impact at very high energy is smaller,
{\it i.e.}~a per cent effect. \s 

\begin{figure}[t!]
\begin{center}
\includegraphics[scale=0.5]{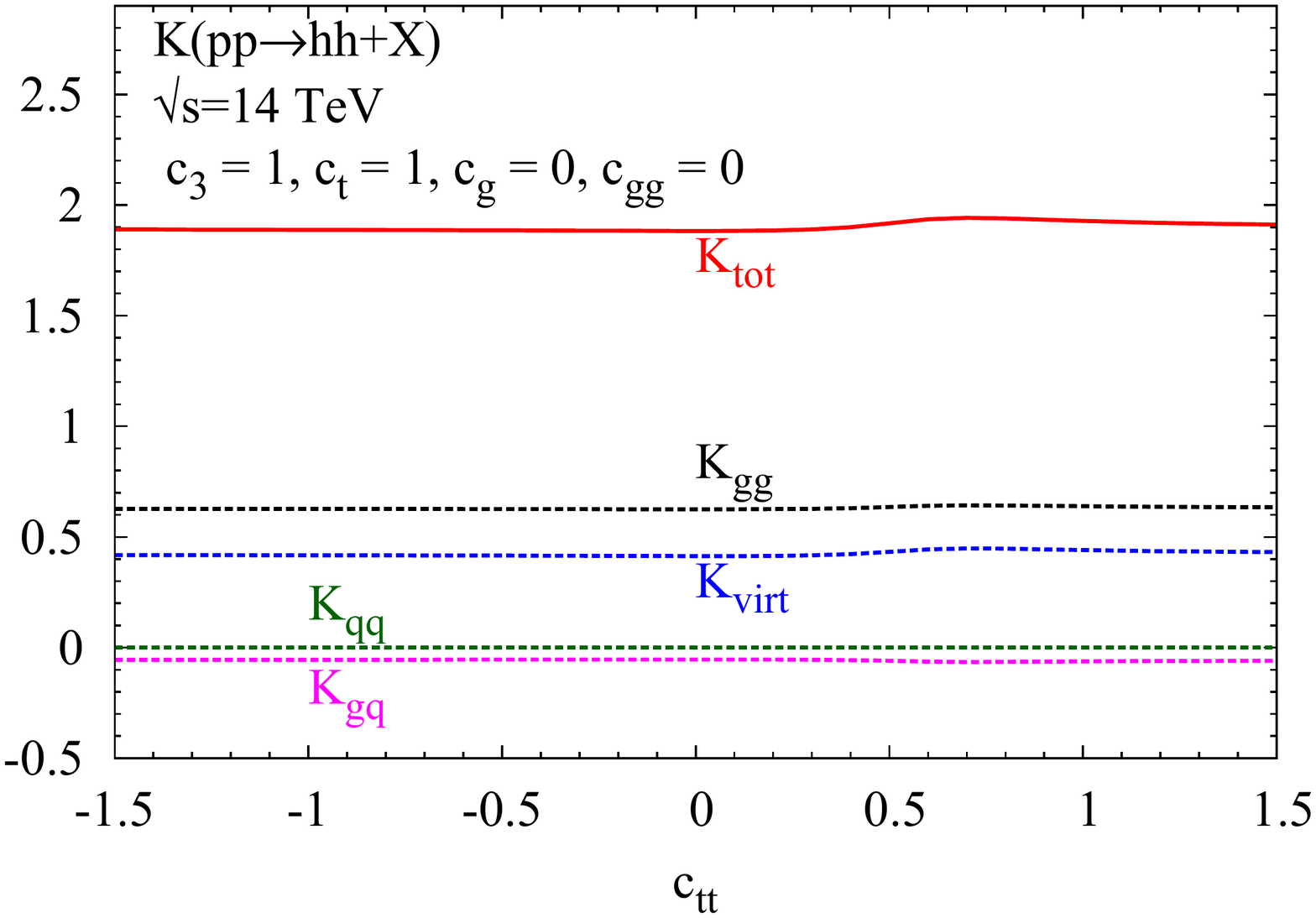}
\vspace*{-1.3cm}
  \caption{Same as Fig.~\ref{fig:kfactorcgandcgg}, but now we have set the SM
    values $c_3=c_t=1$, $c_g=c_{gg}=0$ and varied $c_{tt}$. 
\label{fig:kfactorctt}}
\end{center}
\end{figure}
\begin{figure}[h!]
\begin{center}
\includegraphics[scale=0.5]{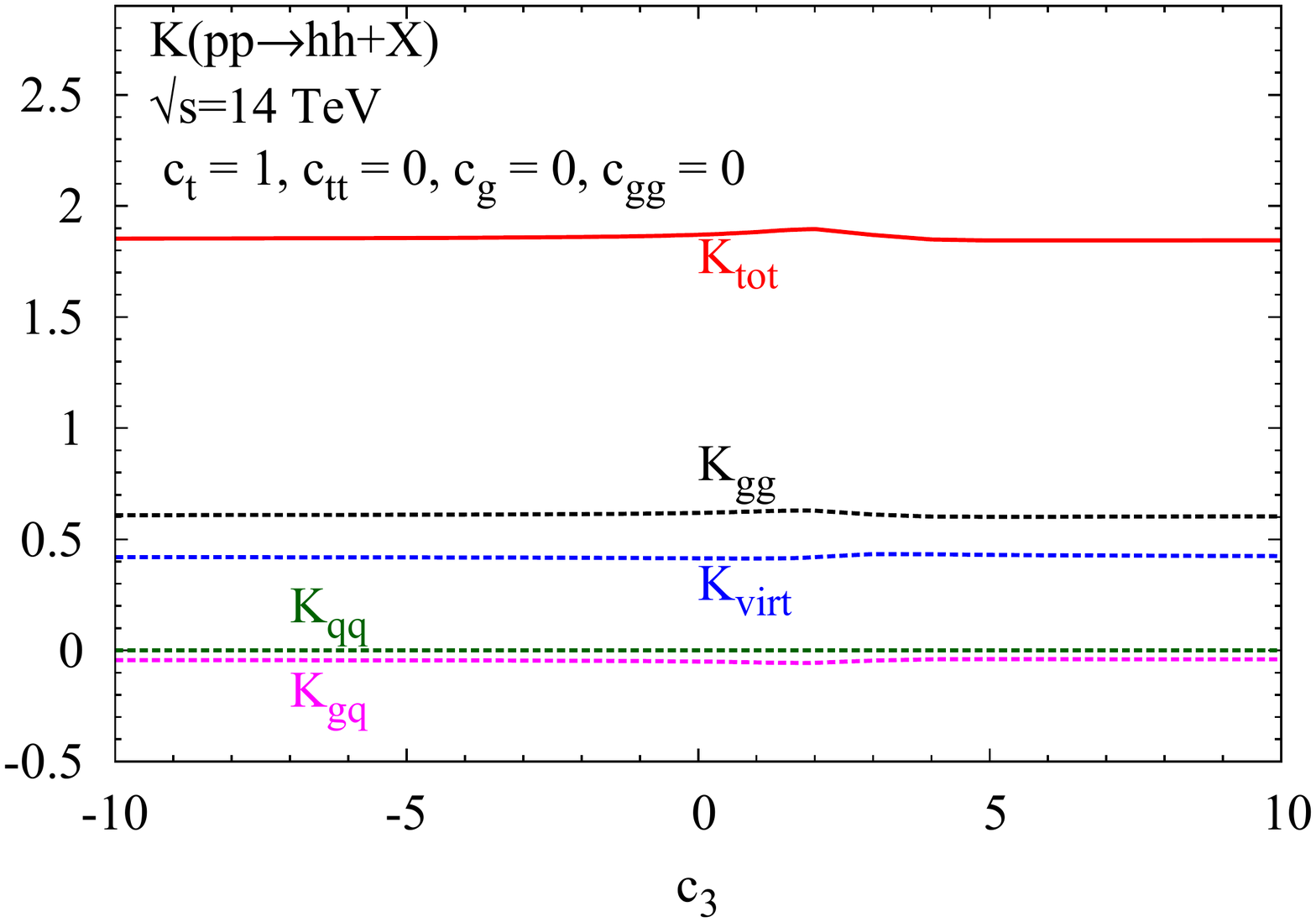}
\vspace*{-1.3cm}
 \caption{Same as Fig.~\ref{fig:kfactorcgandcgg}, but now we have set the SM
   values $c_t=1$, $c_{tt}=c_g=c_{gg}=0$ and varied $c_3$. 
\label{fig:kfactorc3}}
\end{center}
\end{figure} 
In Fig.~\ref{fig:kfactorctt} we have set all couplings to their
SM values but allowed for the new contact interaction between two
Higgs bosons and two top quarks parametrized by $c_{tt}$, which we
have varied between -1.5 and 1.5 \cite{Azatov:2015oxa}. The maximum
$K$-factor is reached for $c_{tt} \approx 0.7$ where the LO cross
section is minimized and we have 
\beq
\delta_{\,\text{max}}^{\,c_{tt}} = 3.2\% \;.
\eeq 
The maximum deviation is smaller than for $c_{gg}$. Note, furthermore,
that the $c_{tt}$ value, for which the deviation is maximal, is
much larger than for the $c_g$ and $c_{gg}$ variations discussed
before. This is due to the normalization of these coupling factors as
can be inferred from the Feynman rules in
Fig.~\ref{fig:effvertices}. At $\sqrt{s}=100$~TeV
the $K$-factor varies between 1.49 and 1.59. \s

The value of the trilinear coupling is practically not constrained and
we have allowed in Fig.~\ref{fig:kfactorc3} for a variation of $c_3$
between -10 and 10 \cite{Azatov:2015oxa}, while setting all other
couplings to their SM values. The impact of $c_3$ 
is rather small. In the investigated range the maximum deviation of
\beq
\delta_{\,\text{max}}^{\,c_3} = 2.1\% \;,
\eeq 
is reached for $5 \lsim c_3 \lsim 10$.
At $\sqrt{s}=100$~TeV the $K$-factor varies between about 1.42 and 1.51.
\s

Finally, the variation of $c_t$ in the still allowed range $0.65 \le
c_t \le 1.15$ \cite{cmshig009}, while keeping all other couplings
at their SM values, hardly changes the $K$-factor, and we do not show the
corresponding results separately. In this parameter configuration $c_t$ enters in the
virtual correction factor $C$, Eq.~(\ref{eq:ccoeffnew}), through the
terms proportional to $c_1$ and $c_2$, and its effect in the nominator almost
cancels against the one in the denominator. \s

The above discussion shows, that the new couplings affect the
$K$-factor by only a few per cent. It has to be kept in mind though
that we varied the couplings only one by one away from their SM
values. The combination of various new couplings could have a 
larger impact on the NLO corrections. 

\section{Conclusions \label{sec:concl}}
We have computed the NLO QCD corrections to Higgs pair production
including dimension-6 operators in the large top mass limit. The
dimension-6 operators lead not only to coupling modifications of the SM Higgs
couplings, but also induce effective gluon couplings to one and two Higgs
bosons and a novel two-Higgs two-top quark coupling. The various
contributions to Higgs pair production are affected differently by the
QCD corrections. Depending on the relative size of the NP
contributions, the $K$-factor is changed by several per cent in the
parameter regions compatible with the LHC Higgs data. 
This small impact on the $K$-factor underlines the dominance of soft
and collinear gluon effects in the QCD corrections. The inclusion of
the QCD corrections in the gluon fusion process based on the effective
theory approach to describe NP, is necessary for reliable predictions
of the cross section. 

\section*{Appendix}
\begin{appendix}
\section{Gluon Fusion into Higgs Pairs in the SILH
  Approximation \label{sec:silhggfus}} 
The SILH approximation of NP effects is valid for small shifts $\delta
\bar{c}_i$ in the Higgs couplings $c_i$ away from the SM values
$c_i^{\text{SM}}$, {\it i.e.}
\beq
\mbox{SILH:} \qquad c_i = c_i^{\text{SM}} (1+ \delta \bar{c}_i) \;,
\qquad \mbox{with} \quad 
\delta \bar{c}_i \ll 1 \;.
\eeq
While in the non-linear case arbitrary values for the coupling
coefficients are allowed and terms quadratic in $\delta c_i$ have to be
taken into account in order to avoid non-physical observables, such as
negative cross sections, in the SILH approach an expansion linear in
$\delta \bar{c}_i$ has to be performed. With
\beq
c_t &=& 1+\delta \bar{c}_t\equiv 1-
\frac{\bar{c}_H+2\bar{c}_u}{2} \;, \qquad
c_{tt} = \delta \bar{c}_{tt} \equiv
-\frac{\bar{c}_H+3\bar{c}_u}{2} \;, \nonumber \\
c_3 &=& 1 + \delta \bar{c}_3 \equiv 1  -
\frac{3\bar{c}_H-2\bar{c}_6}{2} \;, \qquad
c_g = \delta \bar{c}_g = \delta \bar{c}_{gg} \equiv
\bar{c}_g \left( \frac{4\pi}{\alpha_2}\right) \;,
\eeq
{\it cf.}~Eq.~(\ref{eq:nonlincoeff}), this yields for the LO partonic cross
section Eq.~(\ref{eq:losigma}) in the SILH parametrization
\beq
\hat{\sigma}_{\text{LO}}^{\text{SILH}} (gg \to hh) &=&
\hspace*{-0.2cm} \int_{\hat{t}_-}^{\hat{t}_+} 
d\hat{t} \, \frac{G_F^2 \alpha_s^2(\mu_R)}{256 (2\pi)^3} \times 
\label{eq:losigmasilh} \\ && 
\hspace*{-0.2cm} \bigg[
\left| \bar{C}_\Delta F_\Delta + F_\Box \right|^2 +
| G_\Box|^2  + 2\mbox{Re} \bigg\{ \left( \bar{C}_\Delta
  F_\Delta + F_\Box\right) \, \delta \bar{c}_{tt} \, F_\Delta^* \nonumber \\ 
 && \hspace*{-0.2cm} 
+ \left[ |\bar{C}_\Delta F_\Delta|^2 + 3\, 
  \bar{C}_\Delta F_\Delta F_\Box^* + 2 \, (|F_\Box|^2  +
  |G_\Box|^2) \right] \delta \bar{c}_t \nonumber \\
&& \hspace*{-0.2cm}
+ \left(\bar{C}_\Delta F_\Delta + F_\Box \right)^*
\left[ \bar{C}_\Delta F_\Delta \delta \bar{c}_3 + 8 
\left(\bar{C}_\Delta \delta \bar{c}_g + \delta \bar{c}_{gg}
\right) \right] \bigg\} \bigg] \;, \nonumber
\eeq
where
\beq
\bar{C}_\Delta \equiv \lambda_{hhh}^{\text{SM}}
\frac{M_Z^2}{\hat{s}-M_h^2+i M_h \Gamma_h} \;, \qquad \mbox{with}
\qquad
\lambda_{hhh}^{\text{SM}} = \frac{3 M_h^2}{M_Z^2} \;.
\eeq
The NLO SILH cross section is obtained from
Eqs.~(\ref{eq:nlosigma})--(\ref{eq:contrib4}) by replacing 
\beq
\hat{\sigma}_{\text{LO}} \to \hat{\sigma}_{\text{LO}}^{\text{SILH}} \qquad
\mbox{and} \qquad C \to C^{\text{SILH}} \;, 
\eeq
with
\beq
C^{\text{SILH}} &=& \pi^2 + \frac{33-2N_F}{6} \log \frac{\mu_R^2}{Q^2} +
\frac{11}{2} + \left[ \int_{\hat{t}_-}^{\hat{t}_+} d\hat{t} \, 
\tilde{\hat{\sigma}}_{\text{LO}}^{\text{SILH}} \right]^{-1} \times
\nonumber \\
&& 
\mbox{Re} \, \int_{\hat{t}_-}^{\hat{t}_+} d\hat{t}
   \, \bigg\{ \left[c_1 -44 (\bar{C}_\Delta^* \delta \bar{c}_g + 
     \delta \bar{c}_{gg}) \right] (\bar{C}_\Delta F_\Delta + F_\Box )
+  c_1 \left[  F_\Delta \, \delta \bar{c}_{tt} 
\right.
\nonumber  \\
&& \left. 
     + (3
       \bar{C}_\Delta F_\Delta + 4  F_\Box) \delta \bar{c}_t + 8
       (\bar{C}_\Delta +3  F_\Box + 3 \bar{C}_\Delta F_\Delta ) \delta
       \bar{c}_g + 8  \delta \bar{c}_{gg} + \bar{C}_\Delta
       F_\Delta \delta \bar{c}_3 \right]
\nonumber
\\ &&  + \left[ 1 +4 \, \delta \bar{c}_t + 24 \, \delta \bar{c}_g \right] c_2
    \frac{p_T^2}{2 \hat{t} 
      \hat{u}} (Q^2-2M_h^2) G_\Box  \bigg\}
\;, \label{eq:ccoeffsilh}
\eeq
where
\beq
\tilde{\hat{\sigma}}_{\text{LO}}^{\text{SILH}} =
\hat{\sigma}_{\text{LO}}^{\text{SILH}} \left[ \frac{G_F^2 \alpha_s^2
    (\mu_R)}{256 (2\pi)^3} \right]^{-1} \;.
\eeq

\end{appendix}

\vspace*{0.5cm}

\end{document}